%Paper: gr-qc/9306011
%From: anr@ibm-3.MPA-Garching.MPG.DE (Alan Rendall)
%Date: Tue, 8 Jun 93 15:55:06 MDT

\magnification=1200
\def\supp{{\rm supp}}
\def\d{\partial}
\def\tr{{\rm tr}}
\def\ref#1{\lbrack #1\rbrack}
\def\next{\hfil\break\noindent}
\font\title=cmbx12

{\title \centerline{Cosmic censorship for some}
\centerline{spatially homogeneous cosmological models.}}

\vskip 1cm

\noindent
Alan D. Rendall
\footnote*{Supported in part by the NSF grant PHY90-16733 and research
funds provided by Syracuse University}
\next
Max-Planck-Institut f\"ur Astrophysik
\footnote\dag{Present address}
\next
Karl-Schwarzschild-Str. 1
\next
8046 Garching bei M\"unchen
\next
Germany

\vskip .5cm
\noindent
and

\vskip .5cm
\noindent
Physics Department
\next
Syracuse University
\next
Syracuse NY 13244-1130
\next
USA

\vskip 2cm
\noindent
{\bf Abstract}

The global properties of spatially homogeneous cosmological
models with collisionless matter are studied. It is shown
that as long as the mean curvature of the hypersurfaces
of homogeneity remains finite no singularity can occur in
finite proper time as measured by observers whose worldlines
are orthogonal to these hypersurfaces. Strong cosmic
censorship is then proved for the Bianchi I, Bianchi IX and
Kantowski-Sachs symmetry classes.

\vfil\eject

\noindent
{\bf 1.Introduction}

The strong cosmic censorship hypothesis says that the maximal
Cauchy development of generic initial data for the Einstein
equations should be inextendible. Since the maximal Cauchy
development is the largest region of spacetime which is
uniquely determined by initial data, this is essentially the
statement that the time evolution of a spacetime can generically be
fixed by giving initial data. The question of the validity of
strong cosmic censorship is thus the question of predictability
in general relativity. The above definition is in fact only
one partial formalisation of the general idea of strong cosmic
censorship and is based on the suggestion of Eardley and
Moncrief[5] that the right way of looking at the problem is
as a question about global existence of solutions of the
Einstein equations. That this is so will be taken for granted
in what follows.

In any study of cosmic censorship where solutions of the Einstein
equations are specified by giving initial data the choice of
matter model is very important. The opinion has been expressed
in the past that it is best to restrict to vacuum spacetimes, at
least in the beginning. If cosmological spacetimes are considered
then the direct physical relevance of vacuum models (representing a
completely empty universe) is questionable. Thus from the point of
view of physics the best justification for studying vacuum
cosmological models seems to be the idea that what is learned in
the process may usefully be applied at some later date to non-vacuum
spacetimes. In the situations considered in the following the
vacuum solutions appear to play an exceptional role and so
some caution is required when treating vacuum cosmological models
as indicators of the behaviour to be expected in cases when matter
is present. At this time there are unfortunately no results available
on the global structure of cosmological spacetimes containing matter
comparable in generality to those which have been obtained in the
vacuum case[3,4,8,13].

As soon as non-vacuum spacetimes are considered, certain problems
arise with cosmic censorship. Phenomena like shell-crossing and
shocks lead to singularities (or at least to situations which
in our present state of mathematical sophistication we are
forced to consider as singularities in rigorous analytical
investigations) which intuitively are due purely to the matter
itself and have little to do with gravitation. These may
obscure the gravitational singularities which are the true
object of interest in the study of cosmic censorship. These
difficulties have been discussed in detail in [20] where
evidence is presented that a good candidate for a matter
model which avoids these problems is the collisionless gas,
with the dynamics of matter being described by the Vlasov equation.
In the asymptotically flat case a number of things have been
proved about the properties of solutions of the
Einstein equations coupled to matter of this type [14-17,19,21].
In the cosmological case much less is known although a
general theorem concerning the long-time behaviour of a
self-gravitating collisionless gas has been proved in the
context of Newtonian cosmology[18]. The aim of this paper
is to begin investigating the global dynamics of cosmological
solutions of the Vlasov-Einstein system.

The simplest cosmological models are those which are spatially
homogeneous and it is natural to look at these first. They fall
into two broad classes, the Bianchi and Kantowski-Sachs models.
The former can further be subdivided into Bianchi types I-IX.
(For general information on this see [23].)
There is a huge literature concerning the dynamics of homogeneous
cosmological models which generally relies on one of the
following three methods:
\next
a) finding explicit solutions
\next
b) analysing a dynamical system in a small number of variables
\next
c) replacing the full equations by some approximate ones.

\vskip 10pt\noindent
The present work aims to obtain rigorous results concerning
general solutions in a given symmetry class and so methods
a) and c) are ruled out. Furthermore, even in the spatially
homogeneous case the Vlasov-Einstein system involves
infinitely many degrees of freedom (a function of the
momentum variables) and so b) is also not an option here.
A direct approach is necessary. The general idea is to
first obtain the maximal Cauchy development by proving
long-time existence in a constant mean curvature (CMC)
slicing; in the homogeneous case the CMC slices coincide
with the hypersurfaces of homogeneity. Once this has been
done inextendibility can be checked by determining whether
incomplete causal geodesics run into a curvature singularity.

The paper is organised as follows. Section 2 contains some
generalities on Bianchi models and a proof that the
occurrence of singularities in solutions of the
Vlasov-Einstein system with a symmetry of this kind is
controlled by the mean curvature of the hypersurfaces of
homogeneity. A corresponding result for solutions with
Kantowski-Sachs symmetry is obtained in section 3. These
two sections also provide information on the qualitative
behaviour of the mean curvature for the various symmetry
classes. In section 4 it is shown that except in vacuum
any singularity in a solution of the Vlasov-Einstein system
with Kantowski-Sachs or Bianchi IX symmetry must be a
curvature singularity, thus proving strong cosmic censorship
in these two cases. It is also shown that the initial
singularity is a curvature singularity for the other
Bianchi types of Class A, once again under the restriction
that the solution not be vacuum. To prove cosmic censorship
for the remaining Bianchi types of Class A it would be
sufficient to prove the completeness of future-directed
causal geodesics. This is difficult and is carried out only
for the simplest case, Bianchi type I, in the last section.

\vskip .5cm
\noindent
{\bf 2. Bianchi models}

Bianchi models are studied in this paper as a first step
towards understanding more general cosmological solutions of the
Vlasov-Einstein system and so it is appropriate to address
the general question of which
solutions of the Einstein equations should be regarded as
cosmological. A conservative definition, and the one which will be
adopted provisionally here, is that a cosmological spacetime is
one which admits a compact partial Cauchy hypersurface. With this
definition, the only Bianchi types which can occur for a
spatially homogeneous cosmological model are I and IX. However
this does not mean that the other Bianchi types are irrelevant
in the present context because of the existence of locally
spatially homogeneous cosmologies.

\noindent
{\bf Definition} A spacetime is called {\it locally spatially
homogeneous} if each point has a neighbourhood which is
isometric to an open subset of a spatially homogeneous
spacetime.

With this definition, most Bianchi types can be realised in
locally spatially homogeneous spacetimes possessing a compact
Cauchy hypersurface. Discussions of which Bianchi types these are
can be found in [1] and [6]. It can be shown that the universal
cover of a locally spatially homogeneous spacetime is spatially
homogeneous and since passing to the universal cover does not
affect the dynamics it is enough to study the dynamics in the
simply connected case. Then for Bianchi models the spatial
manifold is in a natural way a simply connected 3-dimensional
Lie group. The Kantowski-Sachs case, where passing to the
universal cover brings no special advantage, is considered
separately.

Let $G$ be a simply connected 3-dimensional Lie group and $\{e_i\}$ a
left invariant frame on $G$. Denote the dual coframe by $\{e^i\}$.
Consider a spacetime with underlying manifold $G\times I$, $I$ an
open interval, and a metric of the form $-dt^2+g_{ij}(t)e^i\otimes e^j$.
The initial value problem for the Vlasov-Einstein system will now
be investigated in the case that the metric has this special form
and the distribution function depends only on $t$ and $v^i$, where
$v^i$ denotes the spatial components of the momentum in the frame
$\{e_i\}$. Initial data will be given on the hypersurface
$G\times\{0\}$. The constraints are
$$\eqalignno{
R-k^{ij}k_{ij}+(g^{ij}k_{ij})^2&=16\pi T_{00},&(2.1)       \cr
\nabla^i k_{ij}-\nabla_j(g^{lm}k_{lm})&=-8\pi T_{0j}.&(2.2)}$$
The evolution equations are
$$\eqalignno{
\d_t g_{ij}&=-2k_{ij},&(2.3)                                \cr
\d_t k_{ij}&=R_{ij}+(g^{lm}k_{lm})k_{ij}-2k_{il}k^l_j-8\pi T_{ij}
-4\pi T_{00}g_{ij}+4\pi(g^{lm}T_{lm})g_{ij}.&(2.4)}$$
These equations are written using frame components. The second
fundamental form is denoted by $k_{ij}$. $R$ and $R_{ij}$ are the
Ricci scalar and Ricci tensor of the three-dimensional metric
respectively. The components of the
energy-momentum tensor are denoted by $T_{00}$, $T_{0i}$ and $T_{ij}$.
The explicit form of the latter in terms of the phase space density
$f$ of particles is
$$\eqalignno{
T_{00}(t)&=\int f(t,v^k)(1+g_{rs}v^rv^s)^{1/2}
(\det g)^{1/2}dv^1dv^2dv^3,&(2.5) \cr
T_{0i}(t)&=\int f(t,v^k)v_i
(\det g)^{1/2}dv^1dv^2dv^3,&(2.6)                     \cr
T_{ij}(t)&=\int f(t,v^k)v_iv_j(1+g_{rs}v^rv^s)^{-1/2}
(\det g)^{1/2}dv^1dv^2dv^3.
&(2.7)}$$
The Vlasov equation has the following form.
$$\d f/\d t+[2k^i_jv^j-(1+g_{rs}v^rv^s)^{-1/2}\gamma^i_{mn}v^mv^n]
\d f/\d v^i=0.\eqno(2.8)$$
The Ricci rotation coefficients $\gamma^i_{jk}$ are defined in terms of
the structure constants $C^i_{jk}$ of the Lie algebra of $G$ by the
following relations.
$$\eqalignno{
\gamma^i_{jk}&={\textstyle {1\over 2}}g^{il}
(C_{ljk}+C_{jlk}+C_{klj}),&(2.9)        \cr
C_{ijk}&=g_{kl}C^l_{ij}.&(2.10)}$$
To have a complete set of equations it is necessary to compute
$R_{ij}$ in terms of $g_{ij}$ but for the moment it is enough
to know that $R_{ij}$ is of the form $(\det g)^{-n}\times$
(polynomial in $g_{ij}$ and $C^i_{jk}$).

A local existence theorem will now be proved under the assumption that
initial data $(g_{ij}^0, k_{ij}^0, f^0)$ are given with $f^0(v)$ a $C^1$
function of compact support. The characteristics of (2.8) are the
solutions $V^i(s,t,v)$ of the equation
$$dV^i/ds=2k^i_jV^j-(1+g_{rs}V^rV^s)^{-1/2}\gamma^i_{mn}V^mV^n
\eqno(2.11)$$
with $V^i(t,t,v^i)=v^i$. To control the determinant of the metric the
following consequence of (2.3) will be used.
$$d/dt (\log\det g)=-2g^{ij}k_{ij}.\eqno(2.12)$$
The mean curvature $g^{ij}k_{ij}$ will be of particular importance
in the following and will be denoted from now on by $H$. Now define an
iteration as follows. Let $f_0(t,v)=f^0(v)$, $g_0(t)=g^0$ and
$k_0(t)=k^0$. If $f_n$, $g_n$ and $k_n$ are given for some $n$
determine $V_{n+1}$ by solving  the characteristic equation (2.11)
with $g$ and $k$ replaced by $g_n$ and $k_n$. Let
$$f_{n+1}(t,v)=f^0(V_{n+1}(0,t,v)).\eqno(2.13)$$
Next define an energy-momentum tensor $T_{n+1,\alpha\beta}$ by
replacing $f$ by $f_{n+1}$ and $g$ by $g_n$ in (2.5)-(2.7). Now
determine $g_{n+1}$ and $k_{n+1}$ by solving the linear ordinary
differential equations which result when $T_{\alpha\beta}$, $g$
and $k$ are replaced by $T_{n+1,\alpha\beta}$, $g_n$ and $k_n$
respectively in the right hand side of (2.3) and (2.4) and $g$
and $k$ by $g_{n+1}$ and $k_{n+1}$ on the left hand side.
Let $[0,T_{n+1})$ be the maximal interval on which $g_{n+1}$ is
positive definite. By induction it can be shown that $f_n$, $g_n$
and $k_n$ are $C^1$ on their domains of definition.

Let $|g|$ be the maximum modulus of any component $g_{ij}$ with a
similar definition for $k_{ij}$. Now suppose that for all $n\le N-1$
the following bounds hold.
$$\left.\eqalign{
|g_n-g^0|&\le A_1        \cr
(\det g_n)^{-1}&\le A_2  \cr
|k_n-k^0|&\le A_3}\right\}\eqno(2.14)$$
Suppose further that $|v|\le A_4$ whenever $f_n(t,v)\ne 0$. Here
$A_1$-$A_4$ are positive constants which are for the moment arbitrary.
The characteristic system (2.11) implies a bound of the
form
$$|v|\le P_0+B_4t\ {\rm whenever}\ f_N(t,v)\ne 0.\eqno(2.15)$$
where $B_4$ depends only on $A_1$-$A_4$. As a consequence
(2.5)-(2.7) imply a bound for $T_{N,\alpha\beta}$ depending only
on $A_1$-$A_4$. Next (2.3) and (2.4) imply bounds of the form
$$\left.\eqalign{
|g_N-g^0|&\le B_1t              \cr
|k_N-k^0|&\le B_3t}\right\}\eqno(2.16)$$
where $B_1$ and $B_3$ depend only on $A_1$-$A_4$. If $A_1$-$A_4$ (and
hence $B_1$) are fixed then the inequalities (2.14)
imply an inequality of the form
$$(\det g_N)^{-1}\le B_2\eqno(2.17)$$
whenever $t\le T$ and $T$ is some positive time depending only on
$A_1$-$A_4$. Now fix $A_1$-$A_4$ in such a way that
$A_2 > (det g^0)^{-1}$ and $A_4 > P_0$. Next reduce the size of
$T$ if necessary so that $B_1T<A_1$, $B_2<A_2$, $B_3T<A_3$ and
$P_0+B_4T<A_4$. Then it follows that all iterates exist on the
interval $[0,T)$ and that $g_n$ and $k_n$ are bounded on that
interval independently of $n$.

Consider next the difference of successive iterates for $n\ge 2$.
$$\eqalign{
&|g_{n+1}(t)-g_n(t)|+|k_{n+1}(t)-k_n(t)|     \cr
&\le C\int_0^t |g_n(s)-g_{n-1}(s)|+|k_n(s)-k_{n-1}(s)|
+\|f_{n+1}(s)-f_n(s)\|_\infty ds.} \eqno(2.18)$$
For the difference of the characteristics there is the estimate
$$|dV_{n+1}/ds-dV_n/ds|\le C( |V_{n+1}-V_n|+|g_n-g_{n-1}|
+|k_n-k_{n-1}|) \eqno(2.19)$$
Let
$$\eqalign{
\alpha_n(t)&=\sup\{ |V_{n+1}-V_n|(s,t,v):0\le s\le t, v\in
\supp f_{n+1}(t) \cup\supp f_n(t)\}         \cr
&\qquad+|g_{n+1}(t)-g_n(t)|+|k_{n+1}(t)-k_n(t)|}\eqno(2.20)$$
Then
$$\|f_{n+1}(t)-f_n(t)\|_\infty\le \|f^0\|_{C^1}\alpha_n(t).
\eqno(2.21)$$
Inequalities (2.19)-(2.21) imply that
$$\alpha_n(t)\le C\int_0^t \alpha_n(s)+\alpha_{n-1}(s) ds\eqno(2.22)$$
Applying Gronwall's inequality to this gives
$$\alpha_n(t)\le C\int_0^t \alpha_{n-1}(s)ds\eqno(2.23)$$
It follows from this that $\alpha_n(t)\le C^{n-2}\|\alpha_2\|_\infty t^{n-2}/
(n-2)!$ so that $\{g_n\}$, $\{k_n\}$ and $\{V_n\}$ are Cauchy sequences on
the time interval $[0,T)$. Denote the limits of these sequences by $g$,
$k$ and $V$ respectively. Going back to (2.3) and (2.4), we see that
$\d_t g_n$ and $\d_t k_n$ are uniformly convergent. Thus
$f(t,v)=f^0(V(0,t,v))$
is $C^1$ and $(g,k,f)$ is a classical solution of equations (2.3)-(2.5) on
the interval $[0,T)$. If two solutions with the same initial data are
given define a quantity $\alpha(t)$ in terms of their difference in the
same way that $\alpha_n(t)$ was defined in terms of the difference of
two iterates. Going through the same steps as above leads to an estimate
of the form $\alpha(t)\le C\int_0^t\alpha(s)ds$. By Gronwall's inequality
this means that $\alpha(t)$ is zero and hence that the two solutions agree.
Thus the solution which has been constructed is uniquely determined by the
initial data. If the initial data satisfy (2.1) and (2.2) then so does
the solution of (2.3)-(2.5). To show this define
$$\eqalignno{
C&=R-k^{ij}k_{ij}+(g^{ij}k_{ij})^2-16\pi T_{00}&(2.24)     \cr
C_i&=\nabla^jk_{ij}+8\pi T_{0i}&(2.25)}$$
and note that (2.3)-(2.5) imply homogeneous first order ordinary differential
equations for $C$ and $C_i$.

A close examination of the above proof shows that the size of $T$ is only
restricted by the quantities:
$$|g^0|, (\det g^0)^{-1}, |k^0|, \|f^0\|_\infty\eqno(2.26)$$
and the diameter of $\supp f^0$. Thus if on some time interval $[0,T_1)$
the quantities $|g|$, $(\det g)^{-1}$, $|k|$, $\|f\|_\infty$ and
${\rm diam}\ \supp f$ are bounded a solution exists on $[t,t+\epsilon)$ for
any $t\in [0,T_1)$ and some $\epsilon$ independent of $t$. It can be
concluded that the original solution can be extended to the larger
interval $[0,T_1+\epsilon)$. The following result has now been proved.

\vskip .25cm
\noindent
{\bf Lemma 2.1} Let $(g^0,k^0,f^0)$ be an initial data set for equations
(2.3), (2.4) and (2.8) which has Bianchi symmetry, satisfies the
constraints (2.1) and (2.2) and is such that $f^0(v)$ is a $C^1$
function of compact support. Then there exists a unique corresponding
maximal $C^1$ solution $(g,k,f)$ on an interval $[0,T^*)$. If
$|g|$, $(\det g)^{-1}$, $|k|$ and ${\rm diam}\ {\supp f}$ are
bounded on $[0,T^*)$ then $T^*=\infty$.
\vskip .25cm

In order to investigate the question of global existence the mean curvature
will be examined.
$$\d_t H=R+H^2-12\pi T_{00}+4\pi g^{ij}T_{ij}\eqno(2.27)$$
This equation can usefully be combined with the Hamiltonian constraint
(2.1). An equation which can be obtained in this way is
$$\d_t H=k^{ij}k_{ij}+4\pi(T_{00}+g^{ij}T_{ij})\eqno(2.28)$$
which shows that $H$ is non-decreasing. Now suppose that $H$ is bounded
on $[0,T^*)$. Then from (2.28)
$$H(t)\ge H(0)+\int_0^t (k^{ij}k_{ij})(t^\prime)dt^\prime.\eqno(2.29)$$
Thus $\int_0^{T^*}(k^{ij}k_{ij})(t)dt <\infty$.

At this point a short interlude on linear algebra is necessary. Let $A$
and $B$ be $n\times n$ symmetric matrices with $A$ positive definite.
For any matrix $C$ the norm is defined as usual by
$$\|C\|=\sup\{ \|Cx\|/\|x\|: x\ne 0\}.\eqno(2.30)$$
It is also possible to define a relative norm by
$$\|B\|_A=\sup\{ \|Bx\|/\|Ax\|: x\ne 0\}.\eqno(2.31)$$
It follows immediately that $\|B\|\le\|B\|_A\|A\|$. The other fact
which is needed here is that
$$\|B\|_A\le\sqrt{\tr(A^{-1}BA^{-1}B)}.\eqno(2.32)$$
This can be proved by noting that there exists a basis
$\{z_1,z_2,z_3\}$ with the property that $Bz_i=\lambda_iAz_i$ for
each $i$. The left hand side of (2.32) is then given by the
maximum modulus of any $\lambda_i$. On the other hand, the
eigenvalues of $A^{-1}BA^{-1}B$ are $\{\lambda_i^2\}$ so that
the right hand side of (2.32) is equal to $(\sum_i\lambda_i^2)^{1/2}$.
Let $\|g\|$ and $\|k\|$ denote the norms of the matrices
with entries $g_{ij}$ and $k_{ij}$ respectively. Let $\|k\|_g$ be the
relative norm of the matrix with entries $k_{ij}$ with respect to the
matrix with entries $g_{ij}$. Then
$$\eqalign{
\|g(t)\|&\le \|g(0)\|+2\int_0^t \|k(s)\| ds           \cr
        &\le \|g(0)\|+2\int_0^t \|k(s)\|_{g(s)}\|g(s)\| ds    \cr
        &\le \|g(0)\|+2\int_0^t (k^{ij}k_{ij})^{1/2}(s)\|g(s)\| ds}.
\eqno(2.33)$$
Applying Gronwall's lemma to (2.33) gives
$$\|g(t)\|\le \|g(0)\|\exp
\left[2\int_0^t (k^{ij}k_{ij})^{1/2}(s) ds\right].\eqno(2.34)
$$
By what has been said above it can be concluded that if $T^*$ is
finite then $|g|$ is bounded on
$[0,T^*)$. Equation (2.12) shows that $(\det g)^{-1}$ is bounded on that
interval. It is known that if $\det g$ and its inverse are bounded the
scalar curvature $R$ is bounded from above[9]. Thus, from (2.1),
$k^{ij}k_{ij}$ is bounded on $[0,T^*)$. Using the inequality
$\|k(s)\|\le (k^{ij}k_{ij})^{1/2}\|g(s)\|$ we see that $|k|$ is bounded.
The boundedness of $|g|$ and $(\det g)^{-1}$ implies that $g$ is uniformly
positive definite on the interval of interest. Thus the solutions of the
characteristic equation are bounded there. These statements add up to a
stronger version of Lemma 2.1.

\vskip .25cm
\noindent
{\bf Lemma 2.2} Suppose that the hypotheses of Lemma 2.1 hold. If $H$ is
bounded on $[0,T^*)$ then $T^*=\infty$.

\vskip .25cm
Lemma 2.2 shows that the long time behaviour of $H$ gives useful
information about the long time behaviour of the solution. Assume now
without loss of generality that $H\le0$ for $t=0$. If this is not true
for a given solution it may be arranged by doing the transformation
$t\mapsto -t$. For all Bianchi types other than IX the scalar
curvature $R$ is non-positive[9]. Hence equation (2.1) implies that
for those types $H$ cannot become zero except in very special
circumstances. More specifically, $H=0$ implies that $T_{00}=0$ (so
that the solution is vacuum), $k_{ij}=0$ and $R=0$. Reference to [9]
shows that for these Bianchi types the only cases where $R=0$ is
possible are types I and ${\rm VII}_0$. Moreover the metrics for
which this condition is satisfied are flat.  Thus the spacetime must be
Minkowski space. To sum up: if the Bianchi type is not IX and the
spacetime is not flat then $H<0$ for all time. Lemma 2.2 then implies
that $T^*=\infty$. Next note that $k^{ij}k_{ij}\ge 1/3(g^{ij}k_{ij})^2$
so that (2.28) gives the inequality
$$\d_t H\ge {\textstyle {1\over 3}} H^2.\eqno(2.35)$$
Comparing the solution of this equation with the solution of $\dot u=1/3 u^2$
with the same initial value shows that if $H$ ever becomes positive then it
diverges to $+\infty$ in finite time. Thus if it were known that $H$
always became positive in type IX the ultimate fate of $H$ would be
known in all cases. That this is the case has been shown by Lin and
Wald[10,11] who proved under certain hypotheses (which are all satisfied
by a solution of the Vlasov-Einstein system) that in type IX the mean
curvature cannot remain negative for an infinite time. If it never becomes
positive then, since it is non-decreasing, it must be identically zero
for all $t$ greater than some $t_0$. However it follows from the equations
that in this case the 3-dimensional metric must be flat, contradicting the
fact that the Bianchi type is IX.

\vskip .5cm
\noindent
{\bf 3.Kantowski-Sachs models}

In terms of Gauss coordinates based on a hypersurface of homogeneity,
the Kantowski-Sachs metrics are of the form:
$$ds^2=-dt^2+a^2 dx^2+b^2(d\theta^2+\sin^2\theta d\phi^2),\eqno(3.1)$$
where the functions $a$ and $b$ depend only on $t$. They are
spherically symmetric. Regular frames invariant under this
symmetry do not exist and so an approach must be used which is different
from that of the previous section. The Vlasov equation will be written
in terms of the locally defined orthonormal frame:
$$\d/\d t,\ a^{-1}\d/\d r,\ b^{-1}\d/\d\theta,\ b^{-1}\csc\theta\d/\d\phi.
\eqno(3.2)$$
When written in terms of this frame
the Vlasov equation for general functions in a geometry of the form
(3.1) involves coefficients which are not regular functions on
spacetime. However for distribution functions which are spherically
symmetric these disappear. In the case of distribution functions
$f$ with full Kantowski-Sachs symmetry, i.e. spherically symmetric
and independent of $x$, the equation takes the form:
$$\d f/\d t-a^{-1}\dot av^1\d f/\d v^1-b^{-1}\dot b(v^2\d f/\d v^2
+v^3\d f/\d v^3)=0.\eqno(3.3)$$
In an orthonormal frame the expressions for the components of the
energy momentum tensor are standard.
$$\eqalign{
&\rho=\int f(1+|v|^2)^{1/2} dv^1dv^2dv^3,         \cr
&p_i=\int f(v^i)^2(1+|v|^2)^{-1/2} dv^1dv^2dv^3.}$$
Here $\rho$ is got by contracting $T_{\alpha\beta}$ twice with $e_0$
and $p_i$ is got by contracting it twice with $e_i$. Equations
(2.1)-(2.4) apply to the present situation provided the indices
are now interpreted as coordinate indices. A local existence
theorem for solutions of the Vlasov-Einstein system with
Kantowski-Sachs symmetry can now be proved by following the pattern of
the proof of Lemma 2.1. The arguments leading up to Lemma 2.2 are
also applicable in the present case. In fact the arguments are more
transparent since the metric and second fundamental form are diagonal.
The result of all this is the following:

\vskip 10pt
\noindent
{\bf Lemma 3.1} Let $(g^0,k^0,f^0)$ be an initial data set for
equations (2.3), (2.4) and (2.8) which has Kantowski-Sachs symmetry,
satisfies the constraints (2.1) and (2.2) and is such that $f^0(v)$
is a $C^1$ function of compact support. Then there exists a unique
corresponding maximal $C^1$ solution $(g,k,f)$ on an interval
$[0,T^*)$. If $H$ is bounded on $[0,T^*)$ then $T^*=\infty$.

\vskip 10pt
In order to determine the global behaviour of $H$ it is enough, as
in the Bianchi case, to find out whether in a solution where the
function $H$ is initially negative it will eventually be positive.
An affirmative answer to this question has been obtained by
Burnett[2]. As before the equation (2.35) then implies that
$H$ goes to infinity in finite time. Thus  all Kantowski-Sachs
solutions of the Vlasov-Einstein system exist only for a finite
time in Gauss coordinates and the mean curvature of the hypersurfaces
of homogeneity in such a solution takes on each real value precisely
once.

\vskip .5cm
\noindent
{\bf 4. Curvature singularities}

For any (not necessarily spatially homogeneous) solution of the
Vlasov-Einstein system the particle current density is defined by
$$N^\alpha=-\int fp^\alpha|\det {}^{(4)}g|^{1/2}/p_0
dp^1dp^2dp^3,\eqno(4.1)$$
where $p^\alpha$ denotes the coordinate components of the
momentum. It satisfies
$$\nabla_{\alpha}N^\alpha=0.\eqno(4.2)$$
Now suppose that some region of spacetime is foliated by compact
spacelike hypersurfaces $S_t$ with normal vector $t^\alpha$. Then
the quantity $\int_{S_t} N^\alpha t_\alpha dV_g$, where $dV_g$ is
the volume element of the induced metric on the hypersurface, is
time independent. In the case of a homogeneous spacetime it seems
reasonable to hope that the integral can be dispensed with. For a
Bianchi model $N^0$ takes the form:
$$N^0=\int f (\det g)^{1/2}dv^1dv^2dv^3.\eqno(4.3)$$
Using this equation together with (2.8) and (2.12), the following
identity can be derived:
$$\d_t((\det g)^{1/2}N^0)=-C^i_{ij}g^{jk}T_{0k}(\det g)^{1/2}.
\eqno(4.4)$$
In the Class A models where, by definition, $C^i_{ij}=0$ the right
hand side of (4.4) vanishes and $(\det g)^{1/2}N^0$ is
independent of time. For a Kantowski-Sachs spacetime (4.3) is
replaced by
$$N^0=\int f dv^1dv^2dv^3\eqno(4.5)$$
and a short computation shows that once more $(\det g)^{1/2}N^0$
is constant in time.

Now consider a solution of the Vlasov-Einstein system of the type
discussed in the previous two sections and suppose that for this
solution $H$ tends to infinity in finite time. Let $T^*$ be the
time when $H$ becomes infinite. The proof which was used to show
that this happens in those cases was a comparison with an explicitly
solvable ordinary differential equation. In fact more information
can be extracted from this method. It can be seen that an
inequality of the form $H(t)\ge C(T^*-t)^{-1}$ holds. This inequality
implies that $\det g$ tends to zero as the singularity is approached.
Hence for Kantowski-Sachs and Class A Bianchi models $N^0$ tends to
infinity. Comparing the definitions shows that $T^{00}\ge N^0$ and
so in these cases $T^{00}$ also diverges as the singularity is
approached. Furthermore the scalar $T^{\alpha\beta}T_{\alpha\beta}$
is greater than or equal to $(T^{00})^2$. The Einstein equations then
imply that the curvature invariant $G^{\alpha\beta}G_{\alpha\beta}$
diverges as the singularity is approached. By replacing $t$ by $-t$
this result can also be applied to what is, with the convention for
the time direction chosen in section 2, the initial singularity.
Thus the following result has been proved:

\vskip .25cm
\noindent
{\bf Theorem 4.1} Let $(g,k,f)$ be a non-vacuum locally
spatially homogeneous solution of the Vlasov-Einstein system
with Kantowski-Sachs or Bianchi Class A symmetry which is
the maximal Cauchy development of data on a homogeneous
hypersurface. Then every inextendible past-directed causal
geodesic runs into a curvature singularity. If the symmetry
type is Kantowski-Sachs or Bianchi IX then in addition every
inextendible future-directed causal geodesic runs into a
curvature singularity.

\vskip .25cm
This theorem shows that for Kantowski-Sachs and Bianchi IX
models the maximal Cauchy development of any non-vacuum
initial datum for the Vlasov-Einstein system is inextendible.
Since non-vacuum initial data constitute a generic subset
of all initial data for the Vlasov-Einstein system for any
reasonable definition of genericity, this suffices to prove
strong cosmic censorship within these two symmetry classes.
It is, however, still interesting to ask what happens in the
vacuum case. (When investigating cosmic censorship in
inhomogeneous cosmological models it will be necessary to
consider cases where there is matter somewhere and vacuum
elsewhere.) The only vacuum Kantowski-Sachs model with a
compact Cauchy hypersurface is obtained by identifying the part
of the Schwarzschild solution inside the event horizon by
means of the transformation $t\mapsto t+C$. Here $t$ is the
standard Schwarzschild time coordinate and $C$ a positive
constant. This spacetime has a curvature singularity in the
future but in the past the Schwarzschild event horizon is
now a Cauchy horizon, through which the spacetime can be
extended to a region with closed timelike curves. Thus the
vacuum solution behaves very differently from the solutions
containing matter. In the Bianchi IX case it does not seem
to be known in general which vacuum initial data have
maximal Cauchy developments which are extendible.
However there is an example where the spacetime is
extendible in this way both in the past and the future,
namely the Taub-NUT solution. Siklos[22] has shown that the
Taub-NUT solution is the only only Bianchi type IX vacuum
solution which can be extended to a homogeneous
spacetime containing a Cauchy horizon and so it is
tempting to conjecture that it is the only one which is
extendible beyond the maximal Cauchy development of data
on a homogeneous spacelike hypersurface.

\vskip .5cm
\noindent
{\bf 5. Geodesic completeness}

To investigate the geodesic completeness of homogeneous spacetimes
which expand indefinitely, consider first the following consequence
of (2.27) and the Hamiltonian constraint.
$$\d_t H={\textstyle{ 1\over 4}}(R+H^2+3k^{ij}k_{ij})+4\pi g^{ij}T_{ij}.
\eqno(5.1)$$
This section is concerned with spacetimes of Bianchi type I.
In these spacetimes $R=0$ and so (5.1) implies the inequality
$$\d_t H\ge{\textstyle {1\over 2}}H^2,\eqno(5.2)$$
thus improving on (2.35). Comparing the inequality (5.2) with
the corresponding differential equation gives:
$$H(t)\ge -2(t+C)^{-1},\eqno(5.3)$$
for some constant C. Hence it can be concluded from (2.12) that
$$\det g\le Ct^4.\eqno(5.4)$$
It can also be concluded, using the Hamiltonian constraint, that
$$T_{00}\le (1/4\pi)(t+C)^{-2}.\eqno(5.5)$$
A special feature of Bianchi I spacetimes is that the Vlasov
equation can be solved explicitly. Why this is so and why the
same technique cannot be used for other Bianchi types is explained
in [12]. For Bianchi type I the characteristic equation (2.11)
implies that $dV_i/ds=0$. Thus if the distribution function is
expressed in terms of $v_i$ rather than $v^i$ the Vlasov equation
just says that it is independent of time. Hence
$$T_{00}(t)=\int f(v_i)(1+g^{rs}(t)v_rv_s)^{1/2}(\det g(t))^{-1/2}
dv_1dv_2dv_3.\eqno(5.6)$$
Note that in this equation $f$ does not depend on $t$. The matrix
$g^{ij}(t)$ is symmetric and so there must exist a basis which
is orthonormal with repect to $\delta_{ij}$ and consists of
eigenvectors of this matrix. At any time there must be at least
one of these which corresponds to the largest eigenvalue. Call
it $w_i$. Then in the matrix notation introduced in section 2
$g^{ij}w_iw_j=\|g^{-1}\|$. Moreover
$$g^{ij}v_iv_j\ge g^{rs}w_rw_s(\delta^{ij}v_iw_j)^2
.\eqno(5.7)$$
Assume that $f$ is not identically zero and consider the expression
$$F(y_i)=\int f(v_i)(\delta^{ij}v_iy_j) dv_1dv_2dv_3.\eqno(5.8)$$
If $y_i$ is any vector which lies on the unit sphere in the
sense that $\delta^{ij}y_iy_j=1$ then clearly $F(y_i)$ is
a positive quantity which depends continuously on $y_i$.
Hence it must have a positive minimum $m$ on the sphere and
$m$ depends only on $f$. Putting these facts together leads
to the inequality
$$T_{00}(t)\ge m\|g^{-1}\|^{1/2}(det g)^{-1/2}\eqno(5.9)$$
Using (5.4) and (5.5) in this shows that $\|g^{-1}\|$ is
bounded. Hence each component $g^{ij}$ is bounded and
$g^{ij}V_iV_j$ is bounded along any causal geodesic.

It is obvious that when a homogeneous spacetime described in
Gauss coordinates exists for all $t\ge 0$ then $t$ must tend
to infinity along any inextendible future-directed causal
geodesic. Geodesic completeness can therefore be decided by
looking at the relation between $t$ and an affine
parameter along such a geodesic. They are related by
$$d\tau/dt=(l^2+g^{ij}V_iV_j)^{-1/2}\eqno(5.10)$$
where $l$ is the length of the tangent vector $t^\alpha$
to the geodesic, i.e. $(-t^\alpha t_\alpha)^{1/2}$. Since
$g^{ij}V_iV_j$ is bounded it follows that the right hand side
of (5.10) can be bounded from below by a positive constant.
Hence when (5.10) is integrated to get $\tau$ the integral
diverges as $t\to\infty$ and this is exactly what is
needed to show future geodesic completeness.

\vskip .25cm
\noindent
{\bf Theorem 5.1} Let $(g,k,f)$ be a non-vacuum locally
spatially homogeneous solution of the Vlasov-Einstein system
with Bianchi I symmetry which is the maximal Cauchy
development of data on a spacelike hypersurface. Then every
inextendible future-directed causal geodesic is defined for
arbitrarily large positive values of the affine parameter.

\vskip .25cm

Together with the results of section 4 this proves strong
cosmic censorship in the class of Bianchi I solutions of
the Vlasov-Einstein system.
Once again it is interesting to ask what happens in the
vacuum case. The vacuum equations for Bianchi I spacetimes
can be solved explicitly to obtain the Kasner solutions[23].
It is straightforward to check that these spacetimes
are future geodesically complete and that except in one
special case (which is just a piece of Minkowski space in
disguise) there is a curvature singularity in the past
where the curvature invariant $R^{\alpha\beta\gamma\delta}
R_{\alpha\beta\gamma\delta}$ blows up. In the exceptional
case there is a Cauchy horizon in the past. When the
spatial topology is that of a torus the spacetime can be
extended to a region containing closed timelike curves.
The general vacuum solution has a very different behaviour
at large times from the case where matter is present. It
becomes more and more anisotropic as $t\to\infty$,
contracting in one direction and expanding in two. The
estimates for $g^{ij}$ obtained above show that when
matter is present indefinite contraction in any one
direction is impossible. In this sense the non-vacuum
Bianchi type I solutions of the Vlasov-Einstein system
resemble the Bianchi solutions with dust, which can
be found explicitly[7].

\vskip 10pt
\noindent
{\it Acknowledgements} I thank the Syracuse relativity group
for hospitality during the early stages of this work and in
particular Jorma Louko for various discussions on Bianchi
models.

\vskip .5cm
\noindent
{\bf References}
\next
1.A. ASHTEKAR AND J. SAMUEL, {\it Class. Quantum Grav.} {\bf 8}
(1991), 2191.
\next
2.G. A. BURNETT, {\it Phys. Rev. D} {\bf 43} (1991), 1143.
\next
3.P. CHRU\'SCIEL, {\it Ann. Phys. (N.Y.)} {\bf 202} (1990), 100.
\next
4.P. CHRU\'SCIEL, J. ISENBERG AND V. MONCRIEF, {\it Class. Quantum
Grav.} {\bf 7}(1990), 1671.
\next
5.D. EARDLEY AND V. MONCRIEF, {\it Gen. Rel. Grav.} {\bf 13} (1981),
887.
\next
6.Y. FUJIWARA, H. ISHIHARA AND H. KODAMA, {\it Class. Quantum Grav.}
{\bf 10} (1993), 859.
\next
7.O. HECKMANN AND E. SCH\"UCKING, {\it in} \lq Gravitation: an
Introduction to Current Research\rq\ (L. Witten Ed.) Wiley, New York,
1962.
\next
8.J. ISENBERG AND V. MONCRIEF, {\it Ann. Phys. (N.Y.)} {\bf 199} (1990), 84.
\next
9.R. T. JANTZEN {\it in} \lq Cosmology of the Early Universe,\rq\
(L. Z. Fang and R. Ruffini Eds.) World Scientific, Singapore, 1984.
\next
10.X.-F. LIN AND R. WALD, {\it Phys. Rev. D} {\bf 40} (1989), 3280.
\next
11.X.-F. LIN AND R. WALD, {\it Phys. Rev. D} {\bf 41} (1990), 2444.
\next
12.R. MAARTENS AND S. D. MAHARAJ, {\it Gen. Rel. Grav.} {\bf 22} (1990),
595.
\next
13.V. MONCRIEF, {\it Ann. Phys. (N.Y.)} {\bf 132} (1981), 87.
\next
14.G. REIN, Static solutions of the spherically symmetric
Vlasov-Einstein system, Preprint gr-qc/9304028 (1993).
\next
15.G. REIN AND A. D. RENDALL {\it Commun. Math. Phys.} {\bf 150}
(1992), 561.
\next
16.G. REIN AND A. D. RENDALL {\it Commun. Math. Phys.} {\bf 150}
(1992), 585.
\next
17.G. REIN AND A. D. RENDALL, Smooth static solutions of the
spherically symmetric Vlasov-Einstein system,
{\it Ann. Inst. H. Poincar\'e (Physique Th\'eorique)} (to appear).
\next
18.G. REIN AND A. D. RENDALL, Global existence of classical solutions
to the Vlasov-Poisson system in a three dimensional, cosmological
setting, Preprint (1993).
\next
19.G. REIN, A. D. RENDALL AND J. SCHAEFFER, A regularity theorem for
solutions of the spherically symmetric Vlasov-Einstein system,
Preprint (1993).
\next
20.A. D. RENDALL {\it in} \lq Approaches to Numerical Relativity,\rq\
(R. d'Inverno Ed.) Cambridge University Press, Cambridge, 1992.
\next
21.A. D. RENDALL, The Newtonian limit for asymptotically flat
solutions of the Vlasov-Einstein system,
Preprint MPA 722, gr-qc/9303027 (1993).
\next
22.S. T. C. SIKLOS, {\it Commun. Math. Phys.} {\bf 58} (1978), 255.
\next
23.R. WALD, \lq General Relativity,\rq\ University of Chicago Press,
Chicago, 1984.
\end